\documentstyle[amssymb,aps,prl,twocolumn,epsf]{revtex}

\begin{document}
\draft

\flushbottom
\twocolumn[\hsize\textwidth\columnwidth\hsize\csname
@twocolumnfalse\endcsname
\title{Phonons and Solitons in 1D Mott Insulators}
\author{Chun-Min Chang$^1$, A.~H.~Castro Neto$^{2,1}$, and A.~R.~Bishop$^3$}
\address
{$^1$ Department of Physics,
University of California,
Riverside, CA 92521 \\
$^2$ Department of Physics, 
Boston University, 
Boston, MA 02215\\
$^3$Theoretical Division and Center for Nonlinear Studies, Los
Alamos National Laboratory, Los Alamos, New Mexico 87545}
\date{\today }
\maketitle
\widetext\leftskip=1.9cm\rightskip=1.9cm\nointerlineskip\small
\begin{abstract}
\hspace*{2mm}
We study the problem of one-dimensional (1D) Luttinger liquids in the
insulating Mott-Hubbard phase in the presence of acoustic phonons in the
semiclassical limit. We 
show that solitonic excitations can propagate freely in the system
and the energy required to produce a soliton is reduced by a polaronic
effect. We find a critical value of the electron-phonon coupling
constant for which this energy vanishes.
\end{abstract}
] \narrowtext

It is well-known that phonons can have a major impact in the phase
diagram of 1D interacting electron systems, driving these systems from
charge density waves (CDW) to superconductors due to retardation effects 
\cite{steve}. 
In fact, the problem of phonons in 1D has been studied to a great 
level of detail in the
phase where the linear electronic charge density is incommensurate with the
lattice leading to renormalizations of the various Luttinger liquid parameters
and therefore to no dramatic effect in the charge propagation \cite{martin}. 
In this note we study the problem
of the effect of acoustic phonons in the {\it commensurate} phase of
Luttinger liquids when the 1D system has an instability to the opening of
a gap (Mott gap) in the charge spectrum and therefore to insulating behavior.
In the absence of phonons it can be shown that the lowest energy excitation
is associated with the creation of a soliton (or anti-soliton) 
which can propagate freely
in the system. This excitation, however, requires a finite amount of
energy $\Delta_0$ which is directly related to the Mott gap. 
As it is well-known
the origin of the soliton is associated with the Umklapp scattering
of electrons in the presence of the lattice \cite{thierry}. In a bosonized
description of this process one can show that the bosonic charge 
fields (associated
with electronic charge excitations) are described by a sine-Gordon theory
which has as semi-classical solutions traveling kinks and anti-kinks.
Much less is known about effects of phonons on the soliton propagation
in the insulating phase. We show that while the electron-phonon coupling 
allows for coherent soliton propagation in the Mott phase it
can also 
lead to strong non-perturbative renormalization effects.
We show here 
by a non-perturbative semiclassical 
calculation that the soliton excitation energy is reduced in the 
presence of acoustic phonons and that there is a critical value of the
electron-phonon coupling for which the $\Delta$, the gap for the
soliton creation in the presence of phonons, vanishes. 
Physically, the electron-electron
coupling is suppressed by a polaronic effect.

The basic electronic Hamiltonian we consider 
can be written in a very general form:
\begin{eqnarray}
H_{el}&=&-t\sum\limits_{i,\sigma,\alpha}c_{i,\sigma}^{+}c_{i+1,\sigma
}
+\sum\limits_{i,\sigma,\sigma'} 
V_{\sigma,\sigma'} (i,j)
n_{i,\sigma} \, n_{j,\sigma'} \, ,
\label{Extended Hubbard Model}
\end{eqnarray}
where $c_{i,\sigma}$ ($c^{\dag}_{i,\sigma}$) is the
annihilation (creation) electron operator at lattice site $i$, with spin
projection $\sigma=\uparrow,\downarrow$
and $n_{i,\sigma} = c^{\dag}_{i,\sigma}
c_{i,\sigma}$ is the electron number operator.
Here $t$ is the kinetic energy and $V_{\sigma,\sigma'} (i,j)$ 
is the interaction energy between electrons on different
sites. In what follows we assume short range interactions
only. If, for instance, we keep only the on-site Coulomb
repulsion $V_{\uparrow,\downarrow} (i,i) = U$ 
and the nearest neighbor Coulomb term
$V_{\sigma,\sigma'} (i,i+1) = V$ this Hamiltonian describes the
extended Hubbard Model. The many-body techniques we will apply
here are actually valid for a very generic class of Hamiltonians
\cite{emery}. 
As described elsewhere, \cite{luttinger}
the fermionic degrees of freedom can be written in terms of bosonic charge, 
$\Phi _{\rho }\left( x\right)$, and spin,, 
$\Phi _{\sigma }\left( x\right)$, fields. 
The bosonic version of the fermionic Hamiltonian (\ref{Extended Hubbard Model})
reads:
\begin{eqnarray}
H_{el}&=&H_{\rho }+H_{\sigma }+\frac{2g_{1}}{\left( 2\pi a\right) ^{2}}\int
dx\cos \left[ \sqrt{8}\Phi _{\sigma }\left( x\right) \right] 
\nonumber
\\
&+& \frac{2g_{3}}{%
\left( 2\pi a\right) ^{2}}\int dx\cos \left[ \sqrt{8}\Phi _{\rho }\left(
x\right) +4k_{F}x\right] ,  
\label{bosonized electron hamiltonian}
\end{eqnarray}
where
\begin{eqnarray}
H_{v}=\frac{1}{2}\int dx\left[ \left( u_{v}K_{v}\right) \pi^2 \Pi_{v}^{2}
\left( x\right) +\frac{u_{v}}{K_{v}}\left( \partial_x \Phi
_{v}\left( x\right) \right)^{2} \right] \, ,
\label{bosonic spin or charge separation Hamiltonian}
\end{eqnarray}
where $v=\rho,c$ refers to the 
free bosonic Hamiltonian ($\Pi_v$ is the momentum field operator, that is,
$[\Pi_v(x),\Phi_{u}(y)] = i \delta(x-y) \delta_{u,v}$)
for charge and spin, respectively, 
with Luttinger liquid parameters
$K_{v}$, velocities $u_{v}$, and coupling constants $g_1$ and $g_3$ 
that are directly related to the backscattering and Umklapp processes,
respectively. Here $k_F = \pi n/2$ is the Fermi momentum where $n$ is
the linear density of electrons.
All these parameters
can be written in terms of the bare parameters of the original Hamiltonian
(\ref{Extended Hubbard Model}). In this representation the electronic
charge density can be written as:
\begin{eqnarray}
\rho \left( x\right) =-\frac{\sqrt{2}}{\pi } \partial_x \Phi _{\rho
}\left( x\right).  \label{charge density operator}
\end{eqnarray}
Observe that in (\ref{bosonic spin or charge separation Hamiltonian})
the charge and spin degrees of freedom do not interact with each other.
This is a generic property of 1D quantum fluids (spin-charge separation). 
Many important results
can be understood from the perturbative renormalization group (RG)
studies of (\ref{bosonized electron hamiltonian}). In the presence
of attractive electron-electron interactions the backscattering term, $g_1$,
is a relevant perturbation which leads to the opening of a gap in the
spin spectrum which is associated with the dominance of
superconducting fluctuations. In this case the charge excitations are
gapless. In the presence of repulsive interactions backscattering 
is irrelevant and the Umklapp term, $g_3$, is the most
important one. When the electronic density, $n$, is commensurate with
the lattice (that is, $n=1/a$, where $a$ is the lattice spacing)
the $4 k_F x$ term in (\ref{bosonized electron hamiltonian}) can
be dropped and the Umklapp term becomes a relevant perturbation 
which leads to the opening of a gap in the charge spectrum. In this
case the spin spectrum remains gapless. When $n$ is incommensurate
with the lattice both $g_1$ and $g_3$ are irrelevant and spin and
charge both have gapless spectra. 

Let us now consider the effect of an acoustic phonon field, $\phi(x)$, 
described by the Hamiltonian: 
\begin{eqnarray}
H_{ph}=\frac{1}{2}\int dx\left[ P^{2}\left( x\right)/\rho_s
+c_s^{2} \rho_{s} \left(\partial_x \phi\left( x\right)\right)^{2}\right] ,  
\label{phonon hamiltonian}
\end{eqnarray}
where $P(x)$ is the phonon momentum operator canonically conjugated to
$\phi(x)$, ($[P(x),\phi(y)]=i \delta(x-y)$),
$\rho _{s}=M/a$ is the lattice mass density and $c_s$ is the speed
of sound. As is well-known, \cite{mahan} acoustic phonons couple to electrons
via a deformation potential term which can be written as: 
\begin{eqnarray}
H_{el-ph} &=&-\gamma \frac{\pi }{\sqrt{2}}\int dx\rho \left( x\right) 
\partial_x \phi \left( x\right) 
\nonumber 
\\
&=&\gamma \int dx \partial_x \Phi _{\rho }\left( x\right)
\partial_x \phi \left( x\right) \, ,
\label{electron-phonon hamiltonian}
\end{eqnarray}
where we have used (\ref{charge density operator}). Here $\gamma$ is
the electron-phonon coupling constant. Naively a perturbative 
RG analysis of (\ref{electron-phonon hamiltonian}) would indicate that the 
coupling is irrelevant since it has two derivatives. 
In fact, as we are going to
show the electron-phonon coupling does not affect the soliton propagation 
(and therefore it is irrelevant from this perspective) 
but we are also going to show that the soliton energies are strongly
renormalized by the polaronic effect generated by this operator 
and can even vanish in the
strong coupling regime. Thus, the irrelevancy of the operator in 
the weak couplig regime does not imply that this operator is not
important in the strong coupling limit.

Notice that the spin part of the Hamiltonian (\ref{bosonic spin or charge 
separation Hamiltonian}) does not couple to the phonons 
we consider and can be dropped.
We focus now entirely on the charge degrees of freedom.
When a charge is injected into the
the Mott insulator it creates a soliton excitation which can be described
in terms of the equation of motion for the bosonic fields. Indeed,
consider the equations of motion which are obtained from the Hamiltonian
in the Heisenberg representation. It is very easy to show that the fields
obey the following equations:  
\begin{eqnarray}
\frac{1}{u_{\rho }\pi K_{\rho }}\frac{\partial ^{2}\Phi _{\rho }}{\partial
t^{2}}&=&\frac{u_{\rho }}{\pi K_{\rho }}\frac{\partial ^{2}\Phi _{\rho }}{%
\partial x^{2}}+\frac{\sqrt{2}g_{3}}{\left( \pi a\right) ^{2}}\sin \left( 
\sqrt{8}\Phi _{\rho }\right) +\gamma \frac{\partial ^{2}\phi}{\partial
x^{2}}
\nonumber
\\
\rho _{s}\frac{\partial ^{2}\phi}{\partial t^{2}}&=&c_{s}^{2}\rho _{s}%
\frac{\partial ^{2}\phi}{\partial x^{2}}+\gamma \frac{\partial ^{2}\Phi
_{\rho }}{\partial x^{2}}.
\end{eqnarray}
We now take advantage of the Lorentz invariance of these equations and
make a change of variables 
$\lambda =x \pm \upsilon t$ where $\upsilon $ is the soliton velocity,
in order to write:
\begin{eqnarray}
\left( \frac{\upsilon ^{2}}{u_{\rho }\pi K_{\rho }}-\frac{u_{\rho }}{\pi
K_{\rho }}\right) \frac{\partial ^{2}\Phi _{\rho }}{\partial \lambda ^{2}}&=&%
\frac{\sqrt{2}g_{3}}{\left( \pi a\right) ^{2}}\sin \left( \sqrt{8}\Phi
_{\rho }\right) +\gamma \frac{\partial ^{2}\phi}{\partial \lambda ^{2}}
\nonumber
\\
\rho _{s}\left( \upsilon ^{2}-c_{s}^{2}\right) \frac{\partial ^{2}\phi}{%
\partial \lambda ^{2}}&=&\gamma \frac{\partial ^{2}\Phi _{\rho }}{\partial
\lambda ^{2}}.  
\label{equation of motion for phonon(charge)}
\end{eqnarray}
Combining both equations and replacing $\Phi _{\rho }$\ with $\Phi =\sqrt{8}%
\Phi _{\rho }$\ , the equation of motion for the charge field is given by: 
\begin{equation}
\mu \frac{\partial^2 \Phi}{\partial \lambda ^{2}} +\sin \Phi = 0 \, , 
\label{Equation of motion for charge}
\end{equation}
which is a sine-Gordon equation with
\begin{eqnarray}
\mu =\frac{\left( \pi a\right) ^{2}}{4g_{3}}
\left[ 
\frac{\gamma ^{2}}{\rho_{s} \left( c_{s}^{2}-\upsilon ^{2}\right)}
-\frac{\left(u_{\rho }^{2}-\upsilon ^{2}\right)}{
u_{\rho }\pi K_{\rho }}
\right] \, .   
\label{mu}
\end{eqnarray}
As is well-known this equation has single soliton solutions:
\begin{eqnarray}
\Phi _{\rho }(x,t) &=&\pm \sqrt{2}\tan ^{-1}\left( e^{\frac{1}{\sqrt{\left| \mu
\right| }}\left( x \pm \upsilon t\right) }\right) ,\text{ \ \ \ \ \ \ \ \ \ for }%
\mu >0  \nonumber \\
&=&\pm \sqrt{2}\tan ^{-1}\left( e^{\frac{1}{\sqrt{\left| \mu \right| }}%
\left( x \pm \upsilon t\right) }\right) +\pi ,\text{ \ \ \ for }\mu <0.
\label{solution for charge}
\end{eqnarray}
Moreover, from (\ref{equation of motion for phonon(charge)}), we
find that the lattice deformation also propagates coherently 
and has a profile given by 
\begin{eqnarray}
\phi(x,t) =\pm \frac{\sqrt{2}\gamma }{\rho _{s}\left( \upsilon
^{2}-c_{s}^{2}\right) }\tan ^{-1}\left( e^{\frac{1}{\sqrt{\left| \mu \right| 
}}\left( x \pm \upsilon t\right) }\right) \, .
\label{solution for phonon}
\end{eqnarray}

These equations show that coherent charge propagation is possible 
in the form of a soliton (or kink dressed with a correlated lattice
deformation). The minimum energy required to create
such an excitation
can be obtained by substituting the above solutions for the boson and
phonon field directly into the Hamiltonian. After straightforward
algebra we find that the energy required for soliton creation is given by:
\begin{eqnarray*}
\Delta &=& 
\frac{2}{\pi a}\sqrt{\left| g_{3}\left( \frac{u_{\rho }}{\pi K_{\rho }%
}-\frac{\gamma ^{2}}{\rho _{s}c_{s}^{2}}\right) \right| },
\end{eqnarray*}
which is reduced from its bare value $\Delta_0 = 
\frac{2}{\pi a}\sqrt{g_{3} u_{\rho }/(\pi K_{\rho })}$ by an amount
which depends on the electron-phonon coupling constant. So far this
calculation is semiclassical in nature since it does not consider quantum
corrections which will discussed elsewhere \cite{next}.
This
result predicts, however, that $\Delta$ vanish when $\gamma$
attains a critical value
\begin{eqnarray}
\gamma_c = c_s \sqrt{\frac{u_{\rho} \rho_s}{\pi K_{\rho}}} \, .
\end{eqnarray}
We thus predict that in systems with strong electron-phonon
coupling the energy required to create a soliton vanishes due to the 
polaronic effect described
here. Note that in the absence of the Umklapp term arising from 
the discrete lattice the effects
described here are not possible. 

This effect might explain the recently observed long range charge transport
in DNA (which we claim is a Mott insulator \cite{other}) where
injected charge carriers travel long distances along the DNA double
helix \cite{barton}.

We thank illuminating discussions with J.~Barton, W.~Beyermann, D.~Cox,
G.~Gruner,and S.~Kivelson. We especially thank M.~Pollak for
estimulating our interest in this problem. We acknowledge 
partial support provided by the Collaborative University of California - Los 
Alamos (CULAR)  research grant under the auspices
of the US Department of Energy.

\end{document}